\pdfoutput=1
\documentclass[modern,twocolumn]{aastex631}
\usepackage{amsmath}
\AtBeginDocument{\expandafter\let\csname internallinenumbers\endcsname\relax
  \expandafter\let\csname endinternallinenumbers\endcsname\relax}
\newcommand{\Rs}{R_S}
\newcommand{\avg}[1]{\langle #1 \rangle}

\shorttitle{Switchback growth and the compensated field magnitude}
\shortauthors{Bale, Horbury, and Matteini}

\begin{document}

\title{Radial increase of the compensated interplanetary magnetic field magnitude as a result of the growth of spherically polarized Alfv\'en waves}

\correspondingauthor{Stuart D. Bale}
\email{bale@berkeley.edu}

\author[0000-0002-1989-3596]{S. D. Bale}
\affil{Physics Department, University of California, Berkeley, CA 94720-7300, USA}
\affil{Space Sciences Laboratory, University of California, Berkeley, CA 94720-7450, USA}
\affil{The Blackett Laboratory, Imperial College London, London, SW7 2AZ, UK}

\author[0000-0002-7572-4690]{T. S. Horbury}
\affil{The Blackett Laboratory, Imperial College London, London, SW7 2AZ, UK}
\author[0000-0002-6276-7771]{L. Matteini}
\affil{The Blackett Laboratory, Imperial College London, London, SW7 2AZ, UK}

\begin{abstract}
Measurements from the NASA Parker Solar Probe (PSP) mission demonstrate that
inside $\sim$20 solar radii ($\Rs$) the magnitude of the interplanetary magnetic
field does not fall off simply as $1/R^2$, but rather with $\avg{|B|}\,R^2$ $\sim$20\% below its
value at $\sim$$60\,\Rs$.  Here we use PSP magnetic field measurements over
$10.6$--$56.8\,\Rs$ to show that the outward rise of $\avg{|B|}\,R^2$ is not due to
the global structure of the heliospheric magnetic field or proximity to the
heliospheric current sheet but comes mostly from the growth of large-amplitude,
spherically polarized Alfv\'en waves, whose largest deflections are the so-called `switchbacks',
with a smaller contribution from the Parker spiral.  Over the same
distance range, $\avg{B_R}\,R^2$ stays approximately constant, so that the sector-signed radial
flux is conserved.  The wave amplitude $\epsilon = B_{\perp,{\rm rms}}/\avg{B_\parallel}$
grows from $0.26$ to $0.57$ as $\sim$$R^{1/2}\cos\alpha$, and $\avg{|B|}/\avg{B_R}$ rises
by $20\%$, about $13\%$ from the waves and about $5\%$ from the Parker spiral.  A
simple model parametrized by the broadening of the distribution of
deflection angles reproduces the effect accurately.  The Alfv\'en waves also raise the magnetic pressure in the
low-beta inner heliosphere.  Because $\avg{|B|}$ exceeds $\avg{B_R}$, an open flux estimated from the field
magnitude is too large, which bears on the so-called `open flux problem'.
\end{abstract}

\section{Introduction}

An objective of the NASA Parker Solar Probe (PSP) mission 
has been to measure the radial variation and structure of the interplanetary
magnetic field in the inner heliosphere \citep{2016SSRv..204....7F, 2016SSRv..204...49B}.  Using Encounters 01--05, with perihelia down to approximately $28\,\Rs$,
\citet{2021A&A...650A..18B} found that the open flux estimated by projecting the field onto
the Parker spiral direction, which agrees with the mode of $B_R\,R^2$ close to the Sun, was
approximately independent of radius, with a level near $2.5$ nT~AU$^2$.  Their result supports an approximately $1/R^2$
underlying radial field over that range and motivates testing the scaling
at lower altitudes.

A foundational result from Ulysses is that the scaled radial field
$B_R\,R^2$ is nearly independent of heliographic latitude
\citep{1995Sci...268.1007B, 1995GeoRL..22.3317S, 1996GeoRL..23.3267S}, so that
a single, suitably scaled measurement of $B_R$ estimates the total open flux.  The open flux
measured \emph{in situ} exceeds by a factor of order two the flux that models with a
spherical source surface near $2.5\,\Rs$ \citep{1969SoPh....9..131A,
1969SoPh....6..442S, 1984PhDT.........5H} open from the photosphere, the so-called
`open flux problem' \citep{2017ApJ...848...70L, 2019SoPh..294...19W, 2022SoPh..297...82F}.

The magnetic field in the inner heliosphere is populated by large-amplitude Alfv\'enic deflections,
including switchbacks that reverse its radial component
\citep{2018MNRAS.478.1980H, 2019Natur.576..237B,
2019Natur.576..228K, 2020ApJS..246...39D, 2020ApJS..246...45H}.
These fluctuations complicate the separation of the underlying radial
field from its instantaneous direction \citep{2026SSRv..222...59M}.
See \citet{2026SSRv..222...14B, 2026SSRv..222...43W,
2026SSRv..222...60T, 2026SSRv..222...59M} for recent reviews of their
properties, formation and solar origins.  

Beginning with Encounter 08, with perihelion at $16\,\Rs$, it became apparent that the mode of $B_R\,R^2$ decreases below $\sim$$20\,\Rs$.
Inside $\sim$$20\,\Rs$ both the field magnitude and
the windowed mode of $B_R\,R^2$ appear to fall below the level extrapolated
from larger distances, by 15--20\% at the closest perihelia.  Read as a
measurement of the open flux, this is a departure from $1/R^2$ scaling, and it
would demand a physical explanation near the Sun, such as incomplete opening of the coronal
field, effects due to a non-spherical source surface, reconnection at the heliospheric current
sheet, or a meridional reorganization of the flux.

An alternative explanation lies in the large-amplitude Alfv\'enic fluctuations
themselves and in their evolution as the solar wind expands.
Spherically polarized fluctuations rotate the field at nearly constant
magnitude \citep{1971JGR....76.3534B, 2013LRSP...10....2B, 2015ApJ...802...11M}, and their amplitude relative to the
mean field grows with distance from the Sun \citep{1982JGR....87.3617B, 1990JGR....95.4203R,
2020ApJS..246...53C, 2025arXiv251010106S}.  The mean radial field therefore falls
further and further short of the field magnitude, and $\avg{|B|}\,R^2$ rises outward
even where $\avg{B_R}\,R^2$ stays constant.  In Section~\ref{sec:obs} we describe the
observation.  In Section~\ref{sec:geometry} we describe the geometry that lets the field magnitude and the
radial field fall off differently, and in Section~\ref{sec:results}
we test it against the measurements.

\section{PSP Measurements}
\label{sec:obs}

\begin{figure*}
\begin{centering}
\includegraphics[width=2\columnwidth,clip=true]{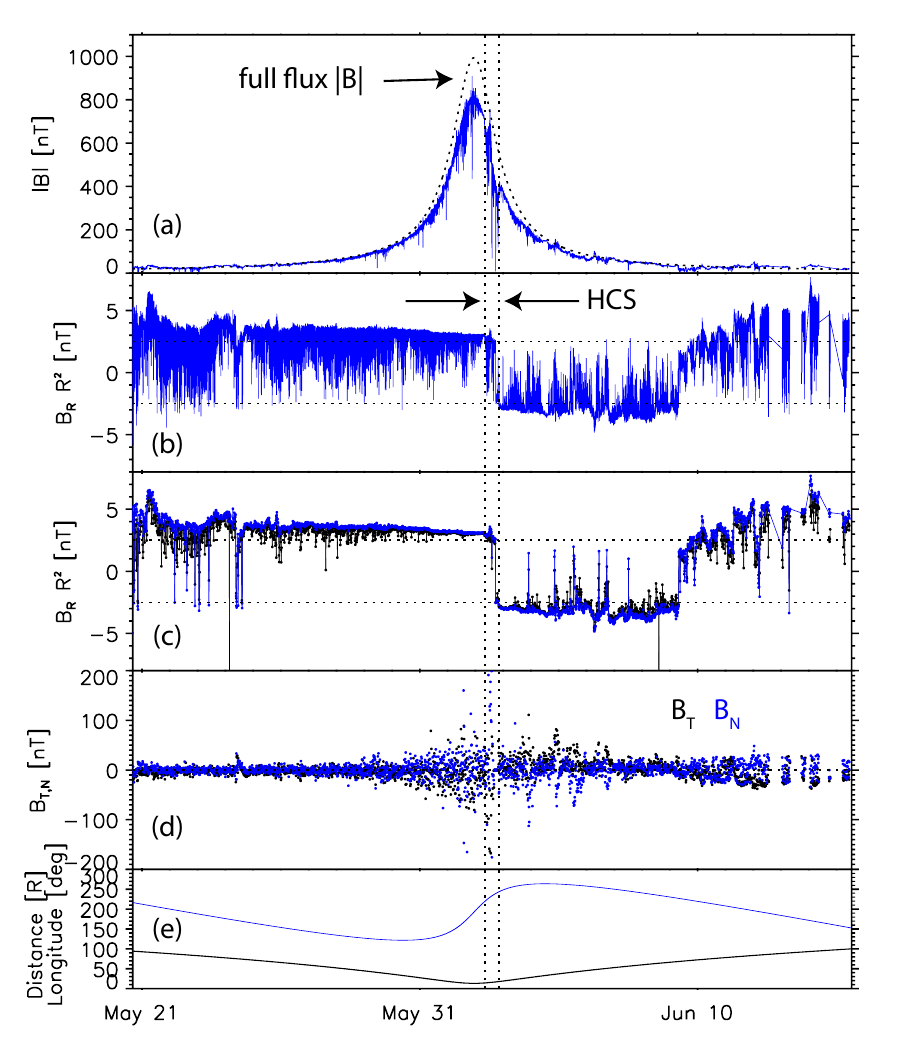}
    \caption{Magnetic field measurements during PSP Encounter 12, with perihelion at
$13.3\,\Rs$ on 2022 June 1 at 23:00 UT.  (a) The field magnitude $|B|$.  The dotted curve
is the value $|B|$ would take if $|B|\,R^2$ held its far-field level.  (b) $B_R\,R^2$, in nT AU$^2$ in panels (b) and (c) with $R$ in AU,
dominated by the large-amplitude Alfv\'enic switchbacks.  Its level decreases near perihelion.  (c) The 15-minute windowed mode (black)
and extreme value (blue) of $B_R\,R^2$, used in earlier work to remove the switchbacks
\citep{2019Natur.576..237B}.  Horizontal dotted lines in (b)
and (c) mark the solar-minimum open flux, 2.5 nT AU$^2$.  (d) The transverse components
$B_T$ (black) and $B_N$ (blue), not scaled by $R^2$, at the times of the extreme values of
$B_R$.  There is no large-scale transverse field, only the residual of switchback
deflections.  (e) Heliocentric distance in $\Rs$ (black) and Carrington longitude (blue).
The interval around the heliospheric current sheet (HCS) crossing is marked and omitted
from the analysis.}
\label{fig:timeseries}
\end{centering}
\end{figure*}

We use fluxgate magnetometer measurements from the FIELDS instrument suite
\citep{2016SSRv..204...49B}, time-averaged to 4 samples per cycle
($\approx 4.58$ Sa/s), from PSP Encounters 01--28.
Figure~\ref{fig:timeseries} shows Encounter 12, with a perihelion of $13.3\,\Rs$ as an example.  Panel (a) shows that $|B|$
peaks near perihelion, and the dotted line marks the value it would take if
$|B|\,R^2$ held the far-field flux level.  Panel (b) shows $B_R\,R^2$, dominated by the switchbacks.  Panel
(d) shows that no coherent large-scale transverse field accompanies the
shortfall in $|B|$.  This excludes a secular $T$ or
$N$ component, which we would expect if the departure of the radial field from
$1/R^2$ were due to some higher-order multipole moment.

\begin{figure*}
\begin{centering}
\includegraphics[width=2\columnwidth,clip=true]{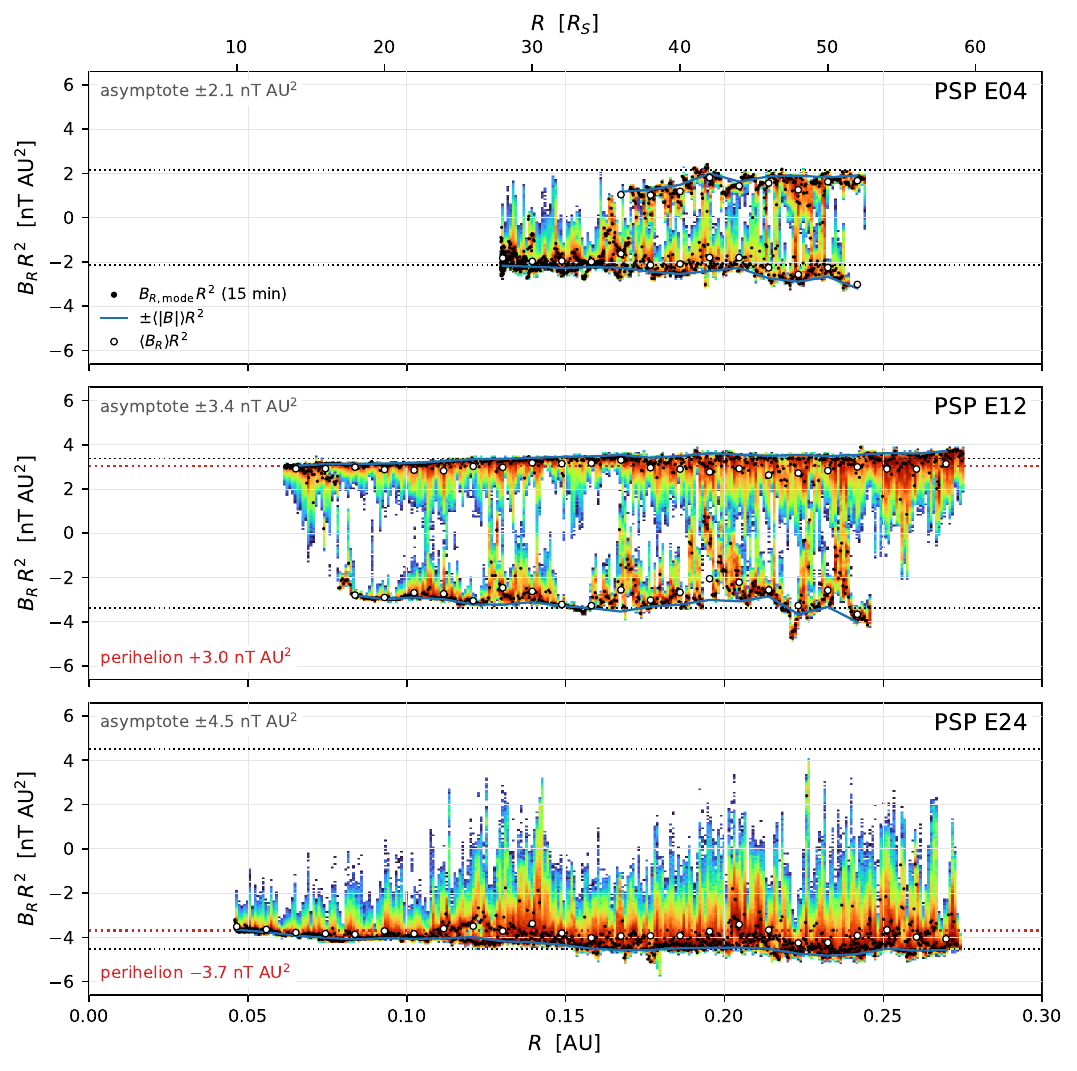}
\caption{Joint distributions of $B_R\,R^2$ and $R$ for PSP Encounters 04, 12
and 24, from every 4 Sa/cycle measurement in bins of $0.18\,\Rs$ by
$0.08$ nT AU$^2$, normalized in each column of $R$ so that every radius
shows its own distribution (logarithmic colour scale).  The
switchbacks appear as the tail that extends from the outer envelope toward zero
and beyond, and the tail broadens outward.  Black dots are the 15-minute windowed
mode of $B_R\,R^2$ following \citet{2019Natur.576..237B}, and the black dotted
lines are its far-field asymptote from a $B_2 + B_3/R$ fit.  In Encounters 12 and
24 the red dotted line marks the near-perihelion level of the mode, the median
within $3\,\Rs$ of perihelion as for $\phi_p$ in Figure~\ref{fig:asymp}, at
$3.0$ and $3.7$ nT AU$^2$ against far-field asymptotes of $3.4$ and $4.5$ nT AU$^2$.
Divided by $R^2$ at perihelion ($13.3$ and $9.9\,\Rs$), the difference is a field
of $\approx 100$ nT in Encounter 12 and $\approx 400$ nT in Encounter 24.
None of these is used as a measurement.  The two curves that
are used, formed in the same windows and binned in $2\,R_S$, are the mean radial field $\avg{B_R}\,R^2$ (white with black edge) and the compensated magnitude
$\pm\avg{|B|}\,R^2$ (blue), signed by sector.  The mean is approximately constant, while the magnitude
tracks the outer envelope and separates from the mean outward.  Encounter 24
never approaches the heliospheric current sheet.}
\label{fig:jointhist}
\end{centering}
\end{figure*}

Figure~\ref{fig:jointhist} shows distributions of $B_R\,R^2$ for Encounters 04, 12, and 24,
with perihelia of $27.9$, $13.3$, and $9.9\,\Rs$, respectively.  These
Encounters sample different phases of the solar cycle.   The windowed mode (black dots) is shown for context.  Figure~\ref{fig:jointhist}
also shows the two quantities we compare below:  the mean radial field
$\avg{B_R}\,R^2$ (white) stays relatively flat, while the compensated field magnitude
$\pm\avg{|B|}\,R^2$ (blue) follows the outer envelope.

\begin{figure*}
\begin{centering}
\includegraphics[width=2\columnwidth,clip=true]{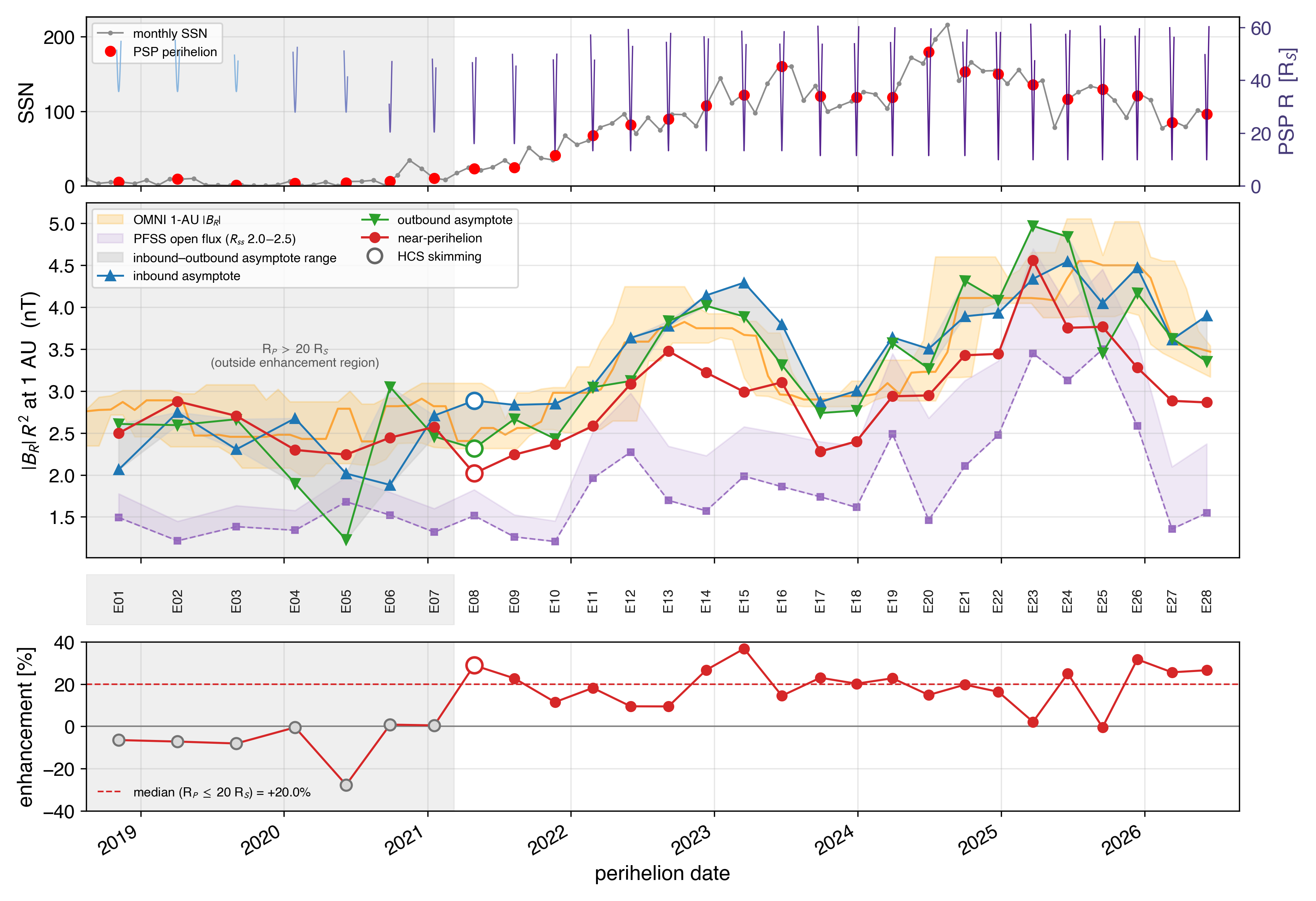}
\caption{The far-field enhancement of the compensated radial field from PSP Encounter 08 onward.  
The top panel shows sunspot number (SSN) and the radial range of PSP for each
Encounter.  The middle panel shows $\phi = |B_R|\,R^2$ from the 15-minute windowed mode
\citep{2019Natur.576..237B} near perihelion (red) and at the inbound (blue) and
outbound (green) asymptotes, with the range between the asymptotes shaded.  The
1 AU OMNI ranges are in orange, and the purple band is a PFSS open-flux estimate
scaled to 1 AU for source-surface heights $R_{ss} = 2.0$--$2.5\,\Rs$.  The mean of the two asymptotes, $\phi_a$, lies above the
near-perihelion value in every Encounter that reaches inside $20\,\Rs$ except Encounter 25, where the two are
nearly equal.  The lower values defined by the PFSS open flux reflect the open flux problem \citep{2017ApJ...848...70L, 2019SoPh..294...19W, 2022SoPh..297...82F}.
The bottom panel shows the fractional enhancement of the far-field asymptote over the
near-perihelion value, $(\phi_a - \phi_p)/\phi_p$.  Encounters 01--07, with
perihelia beyond $20\,\Rs$ (shaded), show no enhancement.  For Encounters
08--28 the median is $+20.0\%$.}
\label{fig:asymp}
\end{centering}
\end{figure*}

Figure~\ref{fig:asymp} shows this effect across PSP Encounters 01--28.  For each Encounter it compares the near-perihelion value
$\phi_p = |B_R|\,R^2$ from the 15-minute windowed mode of
\citet{2019Natur.576..237B} with the mean $\phi_a$ of the inbound and outbound asymptotic values,
with sunspot number for context.  The deficit is absent in Encounters
01--07, whose perihelia lie beyond $20\,\Rs$, and present in every Encounter
from Encounter 08 onward except Encounter 25, where the two values are nearly equal, with a median $(\phi_a - \phi_p)/\phi_p = +20\%$ (bottom panel) over
Encounters 08--28.  The far-field level in the same figure more than
doubles from the 2020 minimum to the 2024--2025 maximum.
In the following analysis, we organize the data in radial `cells' as described in Appendix A.

\section{Magnitude and radial field}
\label{sec:geometry}

We define $\theta_R$ as the angle of the instantaneous field
from radial, so that $B_R = |B|\cos\theta_R$ per sample, and hence
\begin{equation}
  \avg{B_R} = \avg{|B|}\avg{\cos\theta_R} + \mathrm{Cov}\!\left(|B|,\cos\theta_R\right).
  \label{eq:exact}
\end{equation}
Our fluctuations are nearly incompressible and so occur at nearly constant
$|B|$ \citep{1974JGR....79.2302B, 2014GeoRL..41..259M, 2020ApJS..246...32T,
2020ApJ...891L...2S, 2022JPlPh..88e1703S} and expanding-box simulations find that
fluctuations evolve \emph{toward} this state \citep{2024PhPl...31c2901M}.  We measure the rms variation of $|B|\,R^2$ about its mean
$\sigma_{|B|}$ to be $\approx 3\%$ of $\avg{|B|}\,R^2$ (median), so the constant-$|B|$ approximation is
accurate at the few-percent level.  Dividing Equation~(\ref{eq:exact}) by $\avg{|B|}$, the median of
$\mathrm{Cov}(|B|,\cos\theta_R)/\avg{|B|}$ is below $0.1\%$ of $\avg{\cos\theta_R}$ in every
radial bin.  Hence
\begin{equation}
  \frac{\avg{|B|}}{\avg{B_R}} \simeq \frac{1}{\avg{\cos\theta_R}} .
  \label{eq:ratio}
\end{equation}
without assuming a particular form for the distribution of deflections.  

Two effects decrease $\avg{\cos\theta_R}$ with distance:  (a) the mean field lies at an angle
$\alpha$ from radial and (b) the growing Alfv\'enic fluctuations spread the field direction
about that mean.  We measure $\alpha$ as the angle of the mean field direction from radial.
The Parker spiral is its systematic part, although $\alpha$ also includes any
persistent normal component.  Defining $B_\parallel$ as the component along the mean field, we
separate the ratio into two factors,
\begin{equation}
  \frac{\avg{|B|}}{\avg{B_R}} = \frac{\avg{|B|}}{\avg{B_\parallel}}\times\frac{\avg{B_\parallel}}{\avg{B_R}} .
  \label{eq:factor}
\end{equation}
The first factor is the effect of the fluctuations about the mean field.  The second is
$1/\cos\alpha$, the Parker spiral, because $\avg{B_R}=\avg{B_\parallel}\cos\alpha$
(Appendix~\ref{app:data}).

Measured this way between $10.6$ and
$56.8\,\Rs$ the Alfv\'enic fluctuation (first) factor in Equation \ref{eq:factor} rises by about $13\%$, while the
median $\alpha$ grows from about $4^\circ$ to $19^\circ$ and the Parker spiral factor rises by about $5\%$.  The two
multiply, and together give the $\sim$$20\%$ rise of $\avg{|B|}/\avg{B_R}$ in Figure~\ref{fig:results}(c).

Equation~(\ref{eq:factor}) splits the ratio into two factors that we measure directly.  We now
ask whether the fluctuation factor can instead be predicted from one number, the wave amplitude
$\epsilon = B_{\perp,{\rm rms}}/\avg{B_\parallel}$, where $B_\perp$ is the field component
perpendicular to the mean field.  This needs $\avg{\cos\theta}$ in terms of $\epsilon$, where
$\theta$ is the angle of the field from the mean field direction rather than from radial.
We relate the two with a von Mises--Fisher distribution on the sphere,
$p(\mu)\propto e^{\kappa\mu}$ with $\mu=\cos\theta$, where $\kappa$ sets the angular
width (Appendix~\ref{app:vmf}).  Distributions of this kind describe the rotations of
Alfv\'enic solar wind fluctuations at nearly constant $|B|$
\citep{2018ApJ...869L..32M, 2019NCimC..42...16M, 2024ApJ...960L..15B}.  We use it only to
relate these moments, not as a dynamical model of the fluctuations.  The moment of $\cos\theta$ gives
\[
\begin{gathered}
\avg{\cos\theta}=L(\kappa)=\coth\kappa-\kappa^{-1}\\[2pt]
\epsilon^2=\frac{2}{\kappa L(\kappa)}
\end{gathered}
\]
so the measured $\epsilon$ determines $\kappa$.  At small angles
$\kappa \simeq 2/\epsilon^2$ and \citet{2026arXiv261024057L} find the same relation for the
core of the distribution in Encounter 01.  Together the two effects give
\begin{equation}
  \frac{\avg{|B|}}{\avg{B_R}}
  \simeq \frac{1}{L[\kappa(\epsilon)]\cos\alpha}
  \simeq \frac{1+\epsilon^2/2}{\cos\alpha} .
  \label{eq:closure}
\end{equation}
The second approximation is accurate for the values of $\kappa$ we measure
(Appendix~\ref{app:vmf}).  Both $\epsilon(R)$ and $\alpha(R)$ are obtained from the
magnetic field data alone, with no parameter fitted to the ratio, so comparing
Equation~(\ref{eq:closure}) with the measured $\avg{|B|}/\avg{B_R}$ tests whether the geometry
of the fluctuations accounts for its increase.  We use the radial form
$\epsilon\propto R^{1/2}\cos\alpha$ only as an empirical description of the measured amplitude.

\begin{figure*}
\begin{centering}
\includegraphics[width=2.05\columnwidth,clip=true]{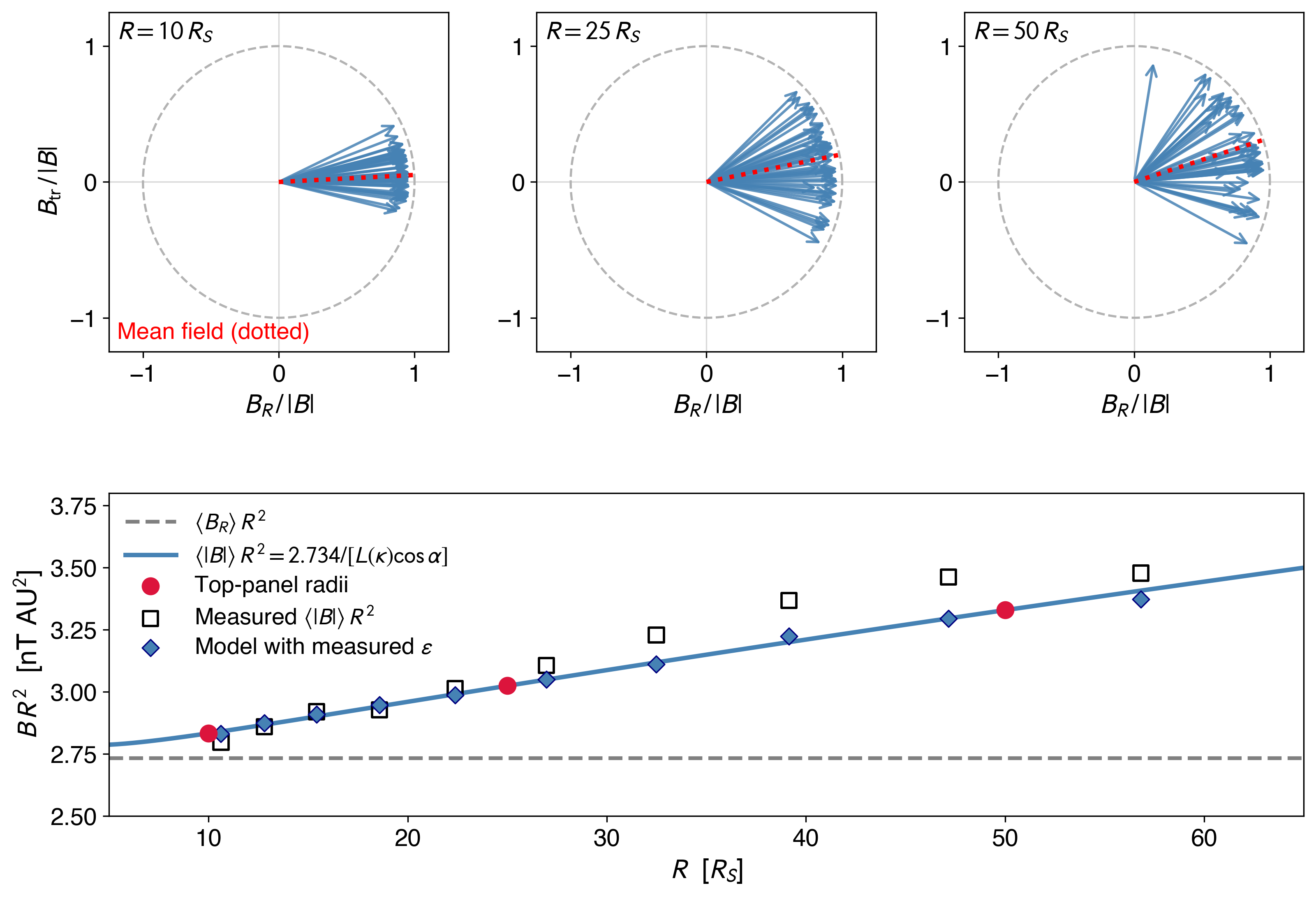}
\caption{Geometric toy model at $R=10$, 25, and $50\Rs$.
Top: projections of 45 sampled unit vectors at each radius onto the plane
containing $\hat{R}$ and $\hat{b}_0$. The
von Mises--Fisher width $\kappa(R)$ is obtained from the fitted
$\epsilon(R)$ in Figure~\ref{fig:results}(b), and the red dotted line marks the
mean-field direction at the Parker spiral angle $\alpha$. The vertical axis is the
signed transverse projection $B_{\mathrm{tr}}/|B|$, along
$\hat{e}_{\mathrm{tr}}=
(\hat{b}_0-\cos\alpha\,\hat{R})/\sin\alpha$.
This component is distinct from the nonnegative $B_\perp$ used in
$\epsilon$. The vectors have unit length in three dimensions, so their
projected tips lie on or inside the unit circle.
Bottom: the analytic closure of Equation~(\ref{eq:closure}),
$\langle|B|\rangle R^2=2.734/[L(\kappa)\cos\alpha]\,\mathrm{nT\,AU^2}$
(blue line), with the compensated mean radial field fixed at
$2.734\,\mathrm{nT\,AU^2}$ (dashed line). Red circles mark the three
illustrated radii. Open squares show the measured compensated-magnitude
bin medians from Figure~\ref{fig:results}(a). Blue diamonds evaluate the same closure
using the measured bin amplitudes instead of the fitted amplitude profile.}
\label{fig:geometry}
\end{centering}
\end{figure*}

Figure~\ref{fig:geometry} shows the fluctuation geometry with a simple toy model, using the fits to the
measured $\alpha(R)$ and $\epsilon(R)$.  At each radius the field has a fixed magnitude, and
its direction fans about a mean field tilted from radial by $\alpha$.  As the fan widens
outward the mean radial projection falls, and $\avg{|B|}\,R^2$ rises above the conserved
radial flux by the factor in Equation~(\ref{eq:closure}).  The model does not propagate the waves or describe individual switchbacks.  It draws random field directions at each radius, with a spread set by the measured $\epsilon(R)$.  Appendix~\ref{app:mc} describes the sampling scheme.

\section{Results}
\label{sec:results}

\begin{figure*}
\begin{centering}
\includegraphics[width=\textwidth,clip=true]{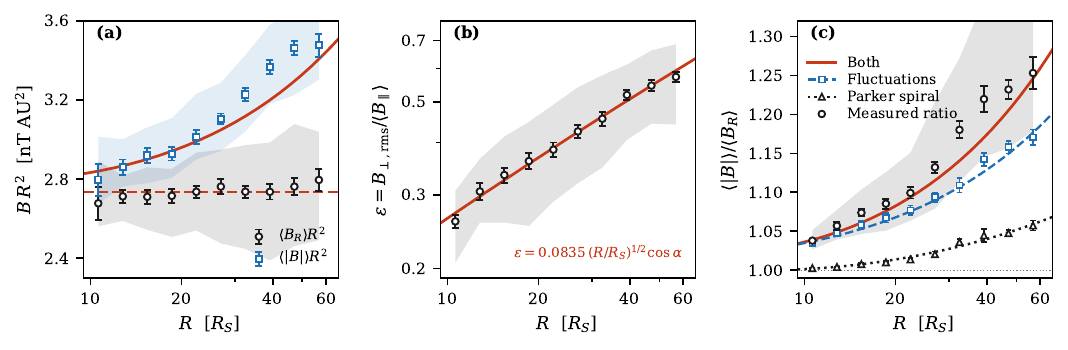}
\caption{Radial profiles of the compensated mean radial field and magnitude, the
fluctuation amplitude, and their ratio.  The data are from Encounters 01--28, in ten
logarithmic bins between $9.7$ and $62.4\,\Rs$.  Symbols are the medians of the cells
in each bin, error bars are bootstrapped over Encounters, and the shading shows the
25--75\% spread of the cells.  (a) $\avg{B_R}\,R^2$ (black) and $\avg{|B|}\,R^2$ (blue),
after one multiplicative offset per Encounter is removed from both.  The red dashed line
is the sample-median flux, $2.73$ nT AU$^2$, and the red curve is that flux times the
closure factor of Equation~(\ref{eq:closure}).  (b) The fluctuation amplitude
$\epsilon = B_{\perp,{\rm rms}}/\avg{B_\parallel}$, measured about the mean field of each
cell, with the fit $\epsilon = 0.0835\,(R/\Rs)^{1/2}\cos\alpha$ in red.  (c) The measured
ratio $\avg{|B|}/\avg{B_R}$ and the closure of Equation~(\ref{eq:closure}) computed from
the fit in (b) (red).  The blue dashed and black dotted curves show its two factors
separately, the fluctuations $1/L(\kappa)$ and the Parker spiral $1/\cos\alpha$.  The blue squares
and black triangles are the same two factors measured directly in each cell,
$\avg{|B|}/\avg{B_\parallel}$ and $\avg{B_\parallel}/\avg{B_R}$, with no model.  The closure and the measured ratio agree to within about $4\%$ in every bin.}
\label{fig:results}
\end{centering}
\end{figure*}

Figure~\ref{fig:results} compares Equation~(\ref{eq:closure}) with the measurements from
Encounters 01--28.  We divide each inbound or outbound leg of an Encounter into \emph{cells} of
60 logarithmic radial bins between $9.2$ and $64.5\,\Rs$, to average over the fluctuations
without averaging over the radial trend.  A cell lasts $\sim$1.4 hours at $10$--$15\,\Rs$ and
$\sim$5.6 hours near $57\,\Rs$.  In each cell we measure $\avg{B_R}\,R^2$ and $\avg{|B|}\,R^2$
on the same samples, so that any difference between them comes from the field direction, and we
measure $\alpha$ and $\epsilon$ about the mean field direction of the cell.  We remove 30 minutes
on either side of each heliospheric current sheet (HCS) crossing, and multiply each field vector
by the sign of its sector, which maps both sectors to an outward field.  We call the result the
sector-signed radial flux.  The radial profiles are medians over the cells in ten logarithmic
bins, after the solar cycle variation (Section~\ref{sec:obs}) is removed by scaling
$\avg{B_R}\,R^2$ and $\avg{|B|}\,R^2$ in each Encounter by the same factor, which leaves their
ratio unchanged.  Appendix~\ref{app:data} gives the details and shows that none of these choices
moves the result by more than 2 percentage points.

The mean field is fitted separately in each of the Encounters with
sufficient radial coverage.  $\avg{B_R}$ falls as $R^{-g}$ with $g=2.00\pm0.02$, as shown in 
Figure~\ref{fig:results}(a).  The radial flux level, $2.73$ nT AU$^2$, is
$9\%$ above the $\sim$$2.5$ nT AU$^2$ that \citet{2021A&A...650A..18B} found over
Encounters 01--05.  

Across the range
in Figure~\ref{fig:results}(a), the magnitude $\avg{|B|}\,R^2$ increases by a factor of $1.24$, compared to
$1.20$ from Equation~(\ref{eq:closure}).  In Figure~\ref{fig:results}(c) the ratio
$\avg{|B|}/\avg{B_R}$ rises by $21\%$ and Equation~(\ref{eq:closure}) by $20\%$.  The Parker spiral contribution lies well below the Alfv\'en wave data.

Comparing radial bins within each Encounter removes the solar cycle.  For each of the 21 Encounters
that span at least $0.9$ e-folds in $R$ (Encounters 08--28), we compare the innermost and
outermost radial bins on the same samples.  Near perihelion $\avg{|B|}\,R^2$ is lower by
$19.1 \pm 2.0\%$ and $\avg{B_R}\,R^2$ by only $2.0 \pm 2.7\%$.  The difference, $17.1 \pm 1.5\%$,
is positive in all 21 Encounters and compares to $15.8\%$ from
Equation~(\ref{eq:closure}).  The near-Sun shortfall is in the field magnitude, not in the
radial flux.

In Figure~\ref{fig:results}(b) $\epsilon$
rises from $0.26$ to $0.57$ and follows $\epsilon=0.0835\,(R/\Rs)^{1/2}\cos\alpha$ to within
$5\%$.

\section{Summary and discussion}
\label{sec:disc}

Our measurements show the following.  (i)~The mean radial field $\avg{B_R}\,R^2$ is approximately constant from
$10.6$ to $56.8\,\Rs$, and any depletion of the sector-signed radial flux over that range is
consistent with zero.  (ii)~On the same samples the compensated magnitude $\avg{|B|}\,R^2$ rises outward,
and the ratio of the two follows $1/\avg{\cos\theta_R}$ cell by cell.  (iii)~At the hours scale of our cells the wave
amplitude $\epsilon$ grows as $R^{1/2}\cos\alpha$.  Equation~(\ref{eq:closure}), which uses
only this amplitude and the Parker spiral angle, reproduces the $20\%$ increase of the ratio
with no parameter fitted to the ratio.  (iv)~The near-Sun `shortfall' of $|B|\,R^2$ reported from PSP is
therefore mostly the growth of the fluctuations, including the switchbacks, not a change in the
sector-signed radial flux.

The total open flux, as distinct from the sector-signed flux, would need an
independent measure of polarity, such as the suprathermal electron strahl
\citep{2013JGRA..118.1868O, 2022SoPh..297...82F}.  A non-spherical source surface
\citep{1978SoPh...60...83S}, reconnection at the HCS, or velocity-shear-driven folding
\citep{2006GeoRL..3314101L, 2020ApJS..246...39D} may contribute a residual of a few percent,
but not the $\sim$20\% effect seen in the field magnitude.

The geometry of the fluctuations also sets the windowed mode of $B_R$ used to recover the underlying field in
earlier work \citep{2019Natur.576..237B, 2021A&A...650A..18B}.  The distribution of
$B_R\,R^2$ piles up against its outer envelope (Figure~\ref{fig:jointhist}), so the mode
follows $|B|$ rather than $\avg{B_R}$ and rises outward with $\avg{|B|}\,R^2$.  This accounts
for the far-field excess in Figure~\ref{fig:asymp}.  An open flux estimated from the
field magnitude or from the mode overstates the sector-signed radial flux, by about $25\%$ at
$57\,\Rs$ for the magnitude.  Using $\avg{B_R}$ instead moves the in situ value toward the
smaller remote estimates and may remove part of the open flux problem
\citep{2017ApJ...848...70L, 2022SoPh..297...82F}.

The fluctuations also raise the magnetic pressure above its mean-field value, by
$1/L(\kappa)^2 \approx 1.4$ at the outer radii, and lower the local $\beta$ by the same factor relative to a mean-field estimate at the same
thermal pressure.  The outward gradient of this wave pressure contributes to the acceleration of
the wind \citep{1971ApJ...168..509B, 1971A&A....13..380A}, and the lower $\beta$ means that the
magnetic field, including the waves, carries more of the pressure balance between neighbouring flux
tubes in the outer corona.

The perpendicular amplitude itself falls as $R^{-1.46\pm0.03}$, more slowly than the
$1/R^2$ of the mean field, and so $\epsilon$ grows.  Note that this is the $R^{-3/2}$
scaling of WKB Alfv\'en waves in a super-Alfv\'enic wind \citep{1980JGR....85.1311H,
1993A&A...270..304V}.  However, our measurements are in a trans-Alfv\'enic regime \citep{2021PhRvL.127y5101K},
where undamped WKB waves would fall off more slowly than we measure.  Indeed, a faster than WKB decay is consistent with
some turbulent dissipation of the fluctuations.  Our results use only the measured $\epsilon$
and do not depend on whether the fluctuations propagate from the Sun or are driven in situ.  The
rate depends on scale.  With Helios, \citet{1982JGR....87.3617B} and \citet{1990JGR....95.4203R}
found that the fluctuation power falls off faster at higher frequencies, and our amplitude shows
the same dependence on the averaging scale.  With a
fixed 15-minute mean field the exponent of $\epsilon/\cos\alpha$ is $\sim$0.27 rather than
$\sim$0.5, because a shorter window assigns more of the rotation to the local mean field, and
\citet{2026arXiv260203724S} find the same for switchback deflection angles.  The ratio
$\avg{|B|}/\avg{B_R}$ does not depend on this choice, and Equation~(\ref{eq:closure}) predicts
the fluctuation factor $\avg{|B|}/\avg{B_\parallel}$ to better than $1\%$ for every window
length we tested.

The switchbacks themselves lie in a heavier tail beyond the smoothly widening von Mises--Fisher
core (Appendix~\ref{app:vmf}), and so they may form an additional population with its own radial
scaling.  \citet{2021ApJ...919L..31T} take switchbacks as deflections beyond $90^\circ$
from the local mean field and find, from PSP to Helios and Ulysses, that their amplitude falls
about as fast as the mean field, faster than the bulk of the fluctuations, which may reflect
saturation on the sphere of constant $|B|$.  The switchbacks still rotate the
field at nearly constant $|B|$, so they enter Equation~(\ref{eq:ratio}) like any other deflection,
and the tail lowers $\avg{\cos\theta}$ by about $1\%$ at most below the von Mises--Fisher value.

Whether the amplitude scaling persists further out can be tested with the outer legs of PSP's
orbit and with Solar Orbiter, and our outermost bins suggest that it may not.  Beyond
${\sim}40\,\Rs$, $\epsilon$ in Figure~\ref{fig:results}(b) grows more slowly than the $R^{1/2}$ fit, and
the measured $\avg{|B|}\,R^2$ in Figure~\ref{fig:geometry} levels off relative to the toy model.
Expanding-box simulations show this kind of saturation once $\delta B/B$ approaches unity.  Then
$\delta B_R$ and $\delta B$ fall off together, and the spread of field directions on the sphere
stops growing \citep{2024PhPl...31c2901M}.  More broadly, the excess of $\avg{|B|}$ over the mean radial field
is geometric, and wherever large-amplitude, nearly constant-magnitude fluctuations ride an
ordered field the magnitude overstates the coherent field.  The same gap appears between the Zeeman
broadening and Zeeman--Doppler imaging fields of cool stars \citep{2012LRSP....9....1R,
2014MNRAS.441.2361V} and between the equipartition and Faraday rotation fields of spiral
galaxies \citep{2015A&ARv..24....4B}, and the geometry described here may be relevant to
reconciling them.

\begin{acknowledgments}
We thank S. T. Badman, B. D. G. Chandran, J. F. Drake, T. Ervin, J. Giacalone,
V. V. Krasnoselskikh, M. G. Linton, T. D. Phan, N. E. Raouafi,
A. A. Schekochihin, N. Sioulas, J. Squire, and M. Velli for valuable discussions.
SDB acknowledges support from the Royal Society Wolfson Visiting Fellowship program.
TSH is supported by STFC grant UKRI1202.
The FIELDS experiment on the Parker Solar Probe spacecraft was designed and
developed under NASA contract NNN06AA01C.  The PSP/FIELDS team acknowledges the
extraordinary contributions of the Parker Solar Probe mission operations and
spacecraft engineering teams at the Johns Hopkins University Applied Physics
Laboratory.
\end{acknowledgments}

\section*{Data Availability}
The Parker Solar Probe FIELDS magnetometer data used in this work are publicly
available from the FIELDS data archive at the University of California, Berkeley
Space Sciences Laboratory (\url{http://fields.ssl.berkeley.edu}). The same data
are archived at the NASA Space Physics Data Facility and are accessible through
CDAWeb (\url{https://cdaweb.gsfc.nasa.gov}).

\facilities{PSP (FIELDS)}

\software{NumPy \citep{2020Natur.585..357H},
          SciPy \citep{2020NatMe..17..261V},
          Matplotlib \citep{2007CSE.....9...90H}}

\appendix

\section{Some data analysis notes}
\label{app:data}

\emph{Sectors and the current sheet.}  The field magnitude is known to decrease near the HCS
\citep{1994JGR....99.6667W, 2020ApJ...894L..19L}, so we remove the crossings.  We assign each
sample a sector polarity from the sign of the 15-minute running mode of $B_R\,R^2$
\citep{2019Natur.576..237B}.  A sign reversal is a sector crossing when it spans at
least two consecutive mode samples (30 minutes).  We exclude samples for which the mode magnitude
is below 20\% of its Encounter median magnitude, and samples within 30 minutes of each crossing.
This identifies 234 crossings.  The median excluded fractions per Encounter are 0.8\% for the
weak-mode criterion and 3\% for the crossing mask.

\emph{Cells.}  A cell comprises the samples from one inbound or outbound Encounter leg within one
of 60 logarithmic radial bins spanning $9.2$--$64.5\,\Rs$ ($\Delta\ln R=0.032$).  We retain 2272 cells containing more than $1\times10^4$ samples, equivalent to approximately 36 minutes at the
nominal cadence ($\approx 4.58$ Sa/s).  In each cell the mean field direction is
$\hat{b}_0 = \avg{\mathbf{B}}/|\avg{\mathbf{B}}|$, formed from the unweighted, sector-signed field
samples, and $\alpha$ is its angle from radial.  The components are
$B_\parallel = \mathbf{B}\cdot\hat{b}_0$ and $B_\perp = (|B|^2-B_\parallel^2)^{1/2}$, and the
wave amplitude is
\[
\epsilon =
\frac{\avg{(B_\perp R^2)^2}^{1/2}}
     {\avg{B_\parallel R^2}}.
\]
The rms variation of $|B|\,R^2$ about its cell mean is $\sigma_{|B|}$, and we accumulate the deflection angle $\theta$ from $\hat{b}_0$ in $0.5^\circ$ bins, and
$\avg{\cos\theta_R}$ and $\avg{\cos^2\theta_R}$ directly.  For unweighted means,
$\avg{B_R} = \avg{B_\parallel}\cos\alpha$ holds exactly because $\hat{b}_0$ is the mean direction
and the mean transverse component vanishes by definition, with no assumption of azimuthal symmetry.
For our $R^2$-weighted moments it holds to $0.11\%$ rms (maximum $2.0\%$).

\emph{Radial profiles.}  The ten logarithmic bins over $9.7$--$62.4\,\Rs$ have
$\Delta\ln R \approx 0.19$, and each holds 82--305 cells from 12--28 separate Encounters.  The
uncertainty on each median comes from a bootstrap over Encounters.  The Encounters are
resampled with replacement 600 times and the bin medians recomputed, and the error bars in
Figure~\ref{fig:results} are the standard deviation of those medians.  The shaded bands are the
25--75\% spread across cells, which is population scatter and not the uncertainty on the median.
The per-Encounter offset is identified against a common power law in $\avg{B_R}\,R^2$.  The smooth
$\alpha(R)$ in Equation~(\ref{eq:closure}) is a fit, linear in $\ln R$, to the bin medians of
$\alpha$.  It gives $\alpha = 3.5^\circ$ at $10.6\,\Rs$ and $19.2^\circ$ at $56.8\,\Rs$, and we use it
only over that radial range, apart from the illustration in Figure~\ref{fig:geometry}.  Fitted separately in each Encounter, the logarithmic slope of
$\avg{B_R}\,R^2$ is $+0.003\pm0.018$ per e-fold.  The exponent of the perpendicular amplitude comes
from a joint logarithmic fit with one offset per Encounter.

\emph{Closure by bin.}  Between $10.6$ and $56.8\,\Rs$ the measured $\avg{|B|}/\avg{B_R}$ rises by
$20.7\%$ and Equation~(\ref{eq:closure}) by $19.9\%$, of which the Parker spiral contributes $5.7\%$
and the fluctuations $13.5\%$.  With the measured $\epsilon$ of each bin in place of the fitted form,
the closure rises by $19.1\%$.  The largest per-bin departure, $4.1\%$ near $39\,\Rs$, comes
mostly from the skew of the cell distributions, because the median of a product is not the product
of the medians.  With each cell's own $\epsilon$ and $\alpha$ the closure holds to about $1\%$ in
every bin.  Across the range $\avg{|B|}\,R^2$ rises by a factor of $1.243$, against $1.199$ from the
closure, and the two agree to within about $5\%$ at every radius.

\emph{Robustness.}  We repeat the analysis with no current sheet mask, with a mask around every
reversal of the 15-minute mode, with 60 or 120 minutes of persistence required, and with minimum
cell sizes of $2\times10^4$ and $4\times10^4$ samples.  Across these variants the rise of
$\avg{|B|}/\avg{B_R}$ lies between $19.1\%$ and $20.9\%$, and its prediction by
Equation~(\ref{eq:closure}) lies between $19.7\%$ and $20.6\%$.  The per-Encounter difference of
Section~\ref{sec:results} stays positive in every Encounter.

\section{The von Mises--Fisher closure}
\label{app:vmf}

The von Mises--Fisher distribution \citep{1953RSPSA.217..295F} is an
axisymmetric distribution on the sphere. For unit vectors
$\hat{b}$ about the mean direction $\hat{b}_0$,
\begin{equation}
p(\hat{b})\,d\Omega=
\frac{\kappa}{4\pi\sinh\kappa}
e^{\kappa\hat{b}\cdot\hat{b}_0}\,d\Omega.\label{eq:vmf}
\end{equation}
Larger $\kappa$ gives a narrower angular distribution. This is the
maximum-entropy distribution for a specified mean vector and also arises
as the steady state of a drift-diffusion model of field direction
\citep{2024ApJ...960L..15B}. Here we use it only as a one-parameter moment closure.
Writing $\mu=\cos\theta=\hat{b}\cdot\hat{b}_0$
and integrating over azimuth gives
\begin{equation}
p(\mu)=\frac{\kappa e^{\kappa\mu}}{2\sinh\kappa},
\qquad -1\leq\mu\leq1.\label{eq:vmfmu}
\end{equation}
The first two moments are
$\langle\cos\theta\rangle=L(\kappa)\equiv\coth\kappa-1/\kappa$
and $\langle\cos^2\theta\rangle=1-2L(\kappa)/\kappa$.
For constant field magnitude, $B_\parallel=|B|\cos\theta$ and
$B_\perp=|B|\sin\theta$, so the amplitude of Section~\ref{sec:geometry} is
\begin{equation}
\epsilon^2\equiv\frac{\langle B_\perp^2\rangle}
{\langle B_\parallel\rangle^2}
=\frac{\langle\sin^2\theta\rangle}{\langle\cos\theta\rangle^2}
=\frac{2}{\kappa L(\kappa)}.\label{eq:epskappa}
\end{equation}
This decreases monotonically with $\kappa$, allowing a unique inversion.
The measured $\epsilon=0.26$--$0.57$ maps approximately to
$\kappa=31$--$7.1$ and $L=0.967$--$0.859$, corresponding within the
closure to rms deflections of roughly $15^\circ$--$31^\circ$.
For $\kappa\to\infty$, the small-angle Gaussian limit gives
$\langle\theta^2\rangle\simeq2/\kappa$. For $\kappa\to0$, the
distribution becomes uniform on the sphere, $L\simeq\kappa/3$, and
$\epsilon\to\infty$ because $\langle B_\parallel\rangle\to0$ while
the perpendicular rms remains finite.

Letting
\[
d\equiv\coth\kappa-1
=\frac{2e^{-2\kappa}}{1-e^{-2\kappa}}.
\]
and using Equation~(\ref{eq:epskappa}), $L=1+d-1/\kappa$ and $\epsilon^2=2/(\kappa L)$ gives
\[
\frac{1}{L}=\frac{1+\epsilon^2/2}{1+d}.
\]
Thus $1+\epsilon^2/2$ overestimates $1/L$ by the relative amount $d$,
which is less than $1.7\times10^{-5}$ for $\kappa\geq5.85$.
Within this closure and parameter range,
\[
\frac{\langle|B|\rangle}{\langle B_\parallel\rangle}
\simeq1+\frac{\epsilon^2}{2}.
\]
For a distribution concentrated at small deflection angles, the same
leading-order relation follows from
$\cos\theta=1-\sin^2\theta/2+O(\theta^4)$, provided the higher angular
moments remain small.  The von Mises--Fisher form fixes the
higher-order terms, and at $\epsilon = 0.57$ they separate
$1/L = 1.162$ from the naive $1/\sqrt{1-\avg{\sin^2\theta}} = 1.147$.  The
closure is therefore insensitive to the assumed distribution at the level of
the leading term and depends on it only through the higher angular moments, which matter most
when large deflections are common.
The measured distribution is close to the von Mises--Fisher form within $\sim$$30^\circ$ of
$\hat{b}_0$ but has a heavier tail, with up to several times the von Mises--Fisher probability
beyond $60^\circ$ and far more beyond $90^\circ$, with the largest relative excess at the innermost radii.  The tail lowers
$\avg{\cos\theta}$ by $0.1$--$1.1\%$ below $L(\kappa)$, which can account for part of the
shortfall of the closure computed from the measured $\epsilon$.  \citet{2026arXiv261024057L}
describe the distribution in Encounter 01 as a mixture of two von Mises--Fisher populations, a
broad one that includes the switchbacks and a field-aligned core.  Log-normal distributions also
carry heavier tails, and they describe solar wind field rotations at kinetic scales
\citep{2015MNRAS.453L..64C} and the field increments in PSP data that contain switchbacks,
increasingly so with distance from the Sun \citep{2024A&A...686A.238L}.  Either form may describe
our deflections better, but we do not pursue that here.

\citet{2019NCimC..42...16M} found flatter distributions in fast wind, $\kappa \approx 1.8$ with
Helios at 0.3 AU ($\sim$$65\,\Rs$) and $\kappa \approx 0.8$ with Ulysses ($\sim$2 AU), from the
rotation angles of the field at large separations.  That measure includes longer scales than our
hours-long cells, and our own amplitude depends on the averaging scale, so the values are not
directly comparable.  Both, however, show the distribution becoming flatter with distance from the
Sun.

\section{The Monte Carlo toy model}
\label{app:mc}

The toy model of Figure~\ref{fig:geometry} draws directions from Equation~(\ref{eq:vmf}) with
$\kappa(R)$ from Equation~(\ref{eq:epskappa}) and nothing else.  The Parker angle is the fit linear in
$\ln R$ to the measured bin medians of $\alpha$, and the amplitude is the fit
$\epsilon = 0.0835\,(R/\Rs)^{1/2}\cos\alpha$ of Figure~\ref{fig:results}(b).  We
invert $\epsilon^2 = 2/[\kappa L(\kappa)]$ numerically for $\kappa(R)$.  At
$R = 10$, $25$ and $50\,\Rs$ this gives $\alpha = 2.9^\circ$, $11.5^\circ$ and $18.0^\circ$, $\epsilon = 0.26$, $0.41$ and $0.56$, and $\kappa = 29.8$, $13.0$ and $7.3$, which correspond to rms deflections of $15^\circ$, $23^\circ$ and
$31^\circ$.  Each vector is drawn by inverse-transform sampling of
Equation~(\ref{eq:vmfmu}).  With $\xi$ uniform on $[0,1)$, the cosine of the
deflection is $\mu = \kappa^{-1}\ln(e^{-\kappa} + 2\xi\sinh\kappa)$, and the
azimuth about $\hat{b}_0$ is drawn independently and uniformly on $[0,2\pi)$.
Each unit vector is then rotated through $\alpha$ into the radial frame.  The top
panels of Figure~\ref{fig:geometry} show 45 such vectors at each radius, drawn
with a fixed random seed.  Only the distribution of $\mu$ enters the ratio.
Because $\hat{b}_0$ is the mean direction of vectors of equal length, the
transverse part of $\avg{B_R}/|B|$ averages to zero for any distribution of
azimuth that keeps $\hat{b}_0$ as the mean direction, and the uniform azimuth affects only the appearance of the top panels.

The lower panel evaluates the analytic closure in Equation~(\ref{eq:closure}) with the compensated
mean radial field fixed at $2.734$ nT AU$^2$, so no parameter is fitted to the magnitude profile.
The Monte Carlo average of $10^6$ vectors at each of the three radii agrees with the analytic angular
factor to better than $0.01\%$.  The unit vectors illustrate direction only.  None of the 135 vectors
in the top panels is reversed, and the von Mises--Fisher distribution puts very little probability on
$B_R < 0$ at any of these radii.

\end{document}